\documentclass[11pt]{article}

\usepackage{amsmath}
\usepackage{amssymb}
\usepackage{amscd}

\def \bx {{\mathbf x}}
\def \by {{\mathbf y}}
\def \bv {{\mathbf v}}
\def \bs {{\mathbf s}}
\def \r {\rho}
\def \NN {{\mathbb N}}

\def \CC {{\mathbb C}}
\def \oon {\frac{1}{n}}
\def \P {{\mathcal P}}

\def \H {{\mathcal H}}

\def \A {{\mathcal A}}
\def \B {{\mathcal B}}
\def \S {{\mathcal S}}
\def \E {{\mathcal E}}
\def \N {{\mathcal N}}

\def \bone {{\mathbf 1}}
\def \bow {{\mathbf 0}}
\def \d {\delta}
\def \s {\sigma}
\def \e {\epsilon}

\title{Propagation of Molecular Chaos by Quantum Systems and the Dynamics of
the Curie-Weiss Model}

\date{ }
\author{Alexander Gottlieb }

\begin{document}
\maketitle

\newtheorem{proposition}{Proposition}
\newtheorem{conjecture}{Conjecture}
\newtheorem{definition}{Definition}
\newtheorem{corollary}{Corollary}
\newtheorem{lemma}{Lemma}
\newtheorem{theorem}{Theorem}

\begin{abstract}
 The propagation of molecular chaos, a tool
of classical kinetic theory, is generalized to apply to quantum
systems of distinguishable particles.
We prove that the Curie-Weiss model of ferromagnetism propagates
molecular chaos and derive the effective dynamics of a single-spin
state in the mean-field limit.  Our treatment differs from the
traditional approach to mean-field spin models in that it concerns the
dynamics of single-particle states instead of the dynamics of infinite-particle
states.
\end{abstract}

\section{Introduction}

The infinite-particle dynamics of spin models with finite-range
interactions --- such as the Ising model --- can be
defined without difficulty in the norm limit of the local (finite-particle)
dynamics \cite{Ruelle}[Section 7.6].   For mean-field spin models
such as the Curie-Weiss model, the infinite-particle dynamics can only
be defined in certain {\it representations} of the infinite-particle
algebra, as the limit in the strong operator topology of the local dynamics
\cite{Emch,BagMor}.
The purpose of this note is to introduce a new approach to the quantum
mean-field dynamics via the propagation of quantum molecular chaos.
The concept of quantum molecular chaos enables us to
comprehend the infinite-particle limit of the Curie-Weiss dynamics
without constructing an infinite-particle dynamics.

For classical mean-field systems, the theory of the propagation of
molecular chaos enables one to study the effective dynamics of finite groups
of particles without defining dynamics of infinite particle states.
We can achieve the same end in the quantum context by utilizing the
analog for quantum systems of theory of the propagation of molecular chaos.
This device is exploited in \cite{NarnSew}, where a quantum version of
propagation of molecular chaos is used to derive the Vlaosov equation from the
dynamics of quantum particles in the continuum.  Their approach
was inspired by \cite{BH}, wherein the Vlasov equation was
derived from the propagation of molecular chaos by mean-field systems of {\it
classical} particles.

   The concept of molecular chaos dates back to Boltzmann~\cite{Boltzmann}.
In order to derive the fundamental equation of the kinetic
theory of gases, Boltzamnn assumed that the molecules of a
nonequilibrium gas were in a state of ``molecular disorder.''
Nowadays, the term ``molecular chaos'' connotes a system of classical
particles that may be regarded as having stochastically independent and
identically distributed random positions and momenta.  The state of
any molecularly chaotic system is characterized by the probability law of a
single particle of the system, and so the temporal evolution of any system
that is at all times in a state of molecular chaos reduces to
that of the probability law of a {\it single}
particle.  Both Boltzmann's equation for dilute gases and Vlasov's equation
for
dilute plasmas may be interpreted as equations that describe the dynamics
of the
position-velocity distribution $f(\bx,\bv)d\bx d\bv$ of a single
particle in a molecularly chaotic gas or plasma.  Because gases and plasmas
remain in a molecularly chaotic state once they have
entered one, i.e., because they ``propagate molecular chaos,''
the kinetic equations of Boltzmann and Vlasov can be thought of as
evolution equations for a single-particle distribution $f(\bx,\bv)d\bx d\bv$.

The concept of {\it propagation} of molecular chaos is due to Kac
\cite{Kac55, Kac}, who
called it ``propagation of the Boltzmann property'' and used it to
derive the homogenous Boltzmann equation from the infinite
particle asymptotics of certain Markovian gas models.
This idea was further developed in work by \cite{Gru, McK75, Uch83, Szn}.
McKean \cite{McK66,McK68} proved the propagation of chaos by systems
of interacting diffusions. See \cite{BH, Sz84, Oelsch,DawsonGartner,
Graham} for more recent work on and some generalizations of  McKean's
propagation of chaos.  For two good surveys of propagation of
chaos and its applications, see  \cite{Sznitman, Meleard}.

This paper is organized as follows.
Quantum molecular chaos is defined in Section \ref{MolecularChaos}, and related
to classical molecular chaos.  Examples of quantum molecular chaos are
provided; it is shown that sequences
of canonical states are often molecularly chaotic.
In Section \ref{PoC} we define the propagation of quantum molecular
chaos.  We then prove that the Curie-Weiss
model propagates molecular chaos and solve the mean-field dynamical equation
for the single-particle state.

\section{Quantum Molecular Chaos}
\label{MolecularChaos}

The definition of molecular chaos current in the probability
literature is equivalent to the following
\cite{Sznitman, Gottlieb}:

\begin{definition}
\label{ClassicalMolecularChaos}
 Let $S$ be a separable metric space.  For each $n
 \in \NN$, let $\r_n$ be a symmetric probability measure on
$S^n$, the $n$-fold Cartesian power of $S$.  (``Symmetric'' means that
the measures of rectangles are invariant under permutations of the
coordinate axes.)   Let $\r$ be a probability measure on $S$.

The sequence $\{\r_n\}$ is $\r$-{\bf chaotic} if the $k$-dimensional
marginal distributions $\r_n^{(k)}$ converge (weakly) to $\r^{\otimes k}$
as $n \longrightarrow \infty$, for each fixed $k \in \NN$.

\end{definition}

The quantum analog of a probability measure
is a state on a C*-algebra with identity.
A {\it state} on a C*-algebra with identity $\A$ is a
positive linear functional on $\A$ that
equals $1$ at the identity element.  The space of states on
$\A$ endowed with the weak* topology will be denoted $\S(\A)$.
Molecular chaos is an
attribute of certain sequences of symmetric probability measures;
we now define {\it quantum} molecular chaos
to be an attribute of certain sequences of symmetric states.

\begin{definition}[Quantum Molecular Chaos]
\label{NoncommutativeChaos}
Let $\A$ be a C*-algebra with identity,
and denote the $n$-th (spatial) tensor power of $\A$
by $\otimes^n \A$.  Let $\r$ be a state on $\A$. For each
$n \in \NN$, let $\r_n$ be a symmetric state on  $\otimes^n \A$, that
is, a state on $\otimes^n \A$ that satisfies
\[  \r_n(A_1 \otimes \cdots \otimes A_n) = \r_n(A_{\pi(1) }
     \otimes \A_{\pi(2)} \otimes \cdots \otimes A_{\pi(n)})
\]
for all permutations $\pi$ of $\{1,2,\ldots,n \}$.
 For each
$k \le n$, let $\r_n^{(k)} \in \S(\otimes^k \A)$ be defined by
\[
  \r_n^{(k)}(B) =
   \r_n (B \otimes {\bf 1}\otimes {\bf 1}\otimes \cdots \otimes {\bf 1}),
\]
for all $B \in \otimes^k \A$,
and let $\r^{\otimes k}$ be defined by the condition that, for all
$A_1,A_2,\ldots,A_k \in \A$,
\[
   \r^{\otimes k} (A_1 \otimes A_2 \otimes \cdots \otimes A_k) =
   \r(A_1) \r(A_2) \cdots \r(A_k).
\]

The sequence $\{\r_n\}$ is $\r$-{\bf chaotic}
if, for each $k \in \NN$, the states $\r_n^{(k)}$ converge weakly* to
$\r^{\otimes k}$ in $\S(\otimes^k \A)$ as $n \longrightarrow \infty$.

The sequence $\{\r_n\}$ is {\bf molecularly chaotic} if it is
$\r$-chaotic for some state $\r$ on $\A$.

\end{definition}

Suppose $\A$ is the algebra generated by the observables for a single
particle of a certain species.   The algebra generated by all
single-particle observables in a system of $n$ {\it distinguishable}
particles of the same species is $\otimes^n \A$, and
states on $\otimes^n \A$ correspond to statistical ensembles of those
$n$-particle systems.  Quantum molecular chaos of a sequence of
$n$-particle states expresses a condition of
quasi-independence of the particles when the number of particles is
very large.

Quantum molecular chaos is related to classical molecular chaos as follows:

\begin{theorem}
\label{Elizabeth}
Let $\A$ be a C* algebra with identity ${\bf 1}$ and for each
$n \in \NN$ let $\r_n$ be a symmetric state on $\otimes^n \A$.
The following are equivalent:
\begin{trivlist}
\item{(i)}
The sequence $\{ \r_n \}$ is $\r$-chaotic in the sense of
Definition~\ref{NoncommutativeChaos}.

\item{(ii)}
For each pair of positive elements $Q_0$ and $Q_1$ satisfying $Q_0 + Q_1 =
{\bf 1}$, the
sequence of probability measures $\{ P_n \}$
 on $\{0,1\}^n$ defined by
\[
   P_n(j_1,j_2,\ldots,j_n) = \r_n(Q_{j_1} \otimes
       Q_{j_2} \otimes \cdots \otimes Q_{j_n}),
\]
is $P$-chaotic in the classical sense of Definition
\ref{ClassicalMolecularChaos},
where $P$ is the probability measure on $\{0,1\}$
defined by
\[
    P(j) = \r(Q_j).
\]

\end{trivlist}
\end{theorem}

\noindent {\bf Proof}:

 It is clear from Definition~\ref{NoncommutativeChaos} that (i)
$\implies$ (ii).   The rest of this proof is devoted to showing that
(i) $\implies$ (ii).

We first establish the following claim:
Let $\P(\S(\A))$ denote the space of regular Borel probability measures on the
state space of $\A$, endowed with the weak* topology as the dual of
$C(\S(\A))$.  Let $\s$ be any state on $\A$.  The measure $\d_{\s}$,
a point mass at $\s$, is is the only measure $\mu \in \P(\S(\A))$ such
that
\begin{equation}
\label{claim}
      \mu \{ \tau \in \S(\A): |\tau(Q) - \s(Q)| \ge \e \} = 0
\end{equation}
for every $\e >0 $ and $Q \in \A$ with $\bow \le Q \le \bone$.   To
prove this claim, first note that (\ref{claim}) holds for {\it every} element
of $\A$ if it holds for those elements $Q$ with $\bow \le Q \le \bone$,
since every element of $\A$ is a linear combination of such positive
elements.  Since (\ref{claim}) holds for every $Q \in \A$, the Borel
measure $\mu$
is supported on arbitrarily small basic open neigborhoods of $\s \in
\S(\A)$, whence it follows that $\mu(\{\s \} ) = 1$,
since $\mu$ is a regular measure.

Next we define a couple of homeomorphisms:
Let $\otimes^{\infty}\A$ denote the inductive limit of the
spatial tensor products $\otimes^n \A$.
A state $\sigma \in \otimes^{\infty}\A $ is
called {\it symmetric} if
\begin{eqnarray*}
 & & \qquad \qquad \sigma(A_1 \otimes A_2 \otimes\cdots \otimes A_n
     \otimes \bone \otimes \bone \otimes \cdots)   \\
& =  &   \sigma(A_{\pi(1)} \otimes A_{\pi(2)} \otimes \cdots \otimes
A_{\pi(k)} \otimes A_{k+1} \otimes \cdots \otimes A_n
\otimes \bone \otimes \bone \otimes \cdots)  \\
\end{eqnarray*}
for all elements $A_1,A_2,\ldots,A_n$ of $\A$,
and all permutations $\pi$ of $\{1,2,\ldots,k\}$, $k \le n$.  Denote
the space of symmetric states on $\otimes^{\infty}\A$ by
$\S_{sym}(\otimes^{\infty}\A)$.
St\o rmer's theorem \cite{Stormer}[Theorem 2.8] states that there exists an
affine homeomorphism $\Phi$ from the space $\P(\S(\A))$ of regular
probability measures on $\S(\A)$ to  $ \S_{sym}(\otimes^{\infty}\A)$, such
that
$\Phi(\delta_{\mu}) = \mu^{\otimes \infty}$.    The classical
precursor of St\o rmer's theorem, de Finetti's theorem \cite{HewSav},
states that
there exists an affine
homeomorphism $\Xi: \P(\P(\{0,1\})) \longrightarrow
\P_{sym}(\{0,1\}^\infty)$ such that $\Xi(\delta_p) = p^{\otimes
\infty}$.

Now assume that condition (ii) holds.

   Let
$\tau$ be an arbitrary but fixed state on $\A$, and for
each $n$, extend $\r_n \in \otimes^n \A$ to the state
\[
    \tilde{\r}_n = \r_n \otimes \tau \otimes \tau \otimes \tau \otimes
\tau \otimes \tau \otimes \cdots
\]
in $\S(\otimes^{\infty}\A)$
Since $\S(\otimes^{\infty}\A)$ is weak*
compact, every subsequence of $\{ \tilde{\r}_n \}$ has cluster points;
condition (ii) will used to prove that
$\r^{\otimes \infty}$ is the {\it only} cluster point of
$\{ \tilde{\r}_n \}$.
It will follow that $\{ \tilde{\r}_n \}$ converges to $\r^{\otimes
\infty}$, which implies that $\{ \r_n\}$ is $\r$-chaotic.

Let $\mu \in \S(\otimes^{\infty}\A)$ be any cluster point of $\{
\tilde{\r}_n \}$, the
limit of the subsequence $\{ \tilde{\r}_{n_k}  \}$.   Because of the
increasing symmetry of the $\tilde{\r}_{n_k}$, the state $\mu$
is symmetric: $\mu \in \S_{sym}(\otimes^{\infty}\A)$.   Suppose that $\mu \ne
\r^{\otimes \infty}$ (this assumption will lead to a contradiction).   Then
$\Phi^{-1}(\mu) \ne \delta_{\r}$ and we have shown that there must exist
${\bf 0} < Q < \bone$ and $\epsilon > 0$ such that
\begin{equation}
\label{contradict1}
 \Phi^{-1}(\mu) \{ \sigma: |\sigma(Q) - \r(Q)| \ge \epsilon \}  > 0.
\end{equation}
Set $Q_0 = Q$ and $Q_1 = \bone - Q$.  Define $P:\S(\A)
\longrightarrow \P(\{0,1\})$,
mapping $\s$ to $ P_{\s}$, by
\[
     P_{\s}(j) = \s(Q_j) \ ; \quad j \in \{0,1\}.
\]
Define $P^{\infty}:\S(\otimes^{\infty}\A)
\longrightarrow \P(\{0,1\}^{\infty})$, mapping $\s $ to $ P^{\infty}_{\s}$, by
\[
  P^{\infty}_{\s}\{ (x_1,x_2,\ldots): x_1 = j_1,
        \ldots, x_n = j_n\} =
        \s(Q_{j_1} \otimes  \cdots \otimes Q_{j_n} \otimes \bone \otimes
        \cdots).
\]
Condition (ii) implies that
$P^{\infty}(\tilde{\r}_{n_k}) \longrightarrow (P_{\r})^{\otimes \infty}$.
Since
$P^{\infty}$ is continuous and  $\tilde{\r}_{n_k}
\longrightarrow \mu$, it follows that
\begin{equation}
\label{contradict2}
P^{\infty}(\mu) = (P_{\r})^{\otimes \infty}.
\end{equation}

   The composite map
$\Xi^{-1} \circ P^{\infty} \circ \Phi$ is affine and continuous, and it maps
$\delta_{\s}$ to $\delta_{P_{\s}}$ for every $\s \in \S(\A)$.
The map $\widetilde{P}:\P(\S(\A)) \longrightarrow
\P(\P(\{0,1\}))$ induced by $P:\S(\A) \longrightarrow \P(\{0,1\})$
is also affine and continuous, and also maps
$\delta_{\s}$ to $\delta_{P_{\s}}$ for every $\s \in \S(\A)$, so
 $\widetilde{P}$ must equal $ \Xi^{-1} \circ P^{\infty} \circ
\Phi$ by the Krein-Milman theorem.  That is, the following diagram commutes:
\[
\begin{CD}
   \P(\S(\A)) @>\Phi>> \S_{sym}(\otimes^{\infty}\A)  \\
       @V{\widetilde{P}}VV            @VV{P^{\infty}}V \\
     \P(\P(\{0,1\})) @>\Xi>> \P_{sym}(\{0,1\}^{\infty})    \\
\end{CD}
\]
By equation (\ref{contradict2}) and the commutativity of the preceding
diagram,
\begin{equation}
\label{contradict3}
\widetilde{P}(\Phi^{-1}(\mu)) = \Xi^{-1}(P^{\infty}(\mu)) =
\delta_{P_{\r}}.
\end{equation}
But equation (\ref{contradict1}) implies that
\[
     \widetilde{P}(\Phi^{-1}(\mu))\{ p \in \P(\{0,1\}):
        |p(0) - P_{\r}(0)| \ge \e \}  > 0,
\]
and this contradicts (\ref{contradict3}).

\noindent $\blacksquare$

\begin{corollary}
Let $\A$ be a C* algebra with identity, and for each
$n \in \NN$ let $\r_n$ be a symmetric state on $\otimes^n \A$.
If $ \r_n^{(2)}$ converges to
$\r \otimes \r$ then $\{ \r_n\} $ is $\r$-chaotic.
\end{corollary}

\noindent {\bf Proof}:

Let $Q_0$ and $Q_1$ be any positive elements of $\A$ such that
$Q_0 + Q_1 = \bone$, and let $P_n$ and $P$ be as in the statement of
Theorem~\ref{Elizabeth}.  The measures $P_n$ are symmetric, and $P_n^{(2)}$
converges to
$P\otimes P$ since $ \r_n^{(2)}$ converges to $\r \otimes \r$.
This suffices to imply that $\{ P_n \}$ is $P$-chaotic
\cite{Sznitman}.  The $\r$-chaos of $\{ \r_n \}$ now follows from
Theorem~\ref{Elizabeth}.

\noindent $\blacksquare$

The following theorem shows that sequences of canonical states for
mean-field systems are often molecularly chaotic:

\begin{theorem}

Let $V$ be an
operator on $\CC^d$ such that $V(x \otimes y) = V(y \otimes x)$ for all
$x,y \in \CC^d$.
Let $V^n_{1,2}$ denote the operator on $\otimes^n \CC^d$ defined by
\[
   V^n_{1,2}(x_1\otimes x_2 \otimes \cdots \otimes x_n) = V(x_1 \otimes
   x_2) \otimes x_3 \otimes \cdots \otimes x_n ,
\]
and for each $i < j \le n$, define $V^n_{ij}$ as acting similarly
on the $i^{th}$ and $j^{th}$ factors of each simple tensor.
Define the states $\r_n \in \S(\otimes^n \CC^d)$ by
\[
\begin{array}{ccc}
   \r_n(A) = \frac{1}{Z} \mathrm{Tr}\left( e^{- H_n} A \right)\ ; & \qquad
      H_n  =  \oon \sum\limits_{i<j} V^n_{ij} \ ; & \qquad
      Z = \mathrm{Tr} \left( e^{- H_n} \right). \\
\end{array}
\]

The sequence $\{\r_n\}$ is $\r$-chaotic if the density operator
for $\r$ is the unique minimizer of the {\it free energy}
\[
     F[D]  = \frac{1}{2} \mathrm{Tr}((D \otimes D) V)  + \mathrm{Tr}(D
                                    \log D)     .
\]
If $V$ is positive definite then $F$ has a unique minimizer.

\end{theorem}
\noindent {\bf Proof Sketch}:
This theorem is the quantum version of Theorems 2 and 4 of
\cite{Spohn}.  The properties of classical entropy that Messer and Spohn used
to prove those theorems are equally true for the von Neumann entropy
$\ -\mathrm{Tr}(D \log D)$ of density operators, at least when $D$
operates on $\CC^d$.
The necessary properties of von Neumann entropy are proved in \cite{LanRob}.

\noindent$\square$

\section{ The Curie-Weiss Model Propagates Chaos}
\label{PoC}

A sequence of $n$-particle dynamics ``propagates chaos''
if molecularly chaotic sequences of initial distributions remain
molecularly chaotic for all time
under the $n$-particle dynamical evolutions.

In the classical context, the $n$-particle dynamics are Markovian.
Accordingly, in \cite{Gottlieb},
we defined propagation of chaos in terms of Markov transition
functions:
\begin{definition}
\label{PropagationOfClassicalChaos}
For each $n \in \NN$, let $K_n:S^n \times \sigma(S^n) \times [0,\infty)
\longrightarrow [0,1]$ be a Markov transition function which commutes
with permutations in the sense that
\[
     K_n(\bx,E,t) = K_n(\pi \cdot \bx, \pi \cdot E,t)
\]
for all permutations $\pi$ of the $n$ coordinates of $\bx$ and the
points of $E \subset S^n$, and for all $t \ge 0$.  (Here, $\sigma(S^n)$
denotes the Borel $\s$-field of $S^n$.)

The sequence $\{K_n\}_{n=1}^{\infty}$ {\bf propagates chaos} if, for
all $t \ge 0$, the molecular chaos of a sequence  $\{ \r_n \}$ entails
the molecular chaos of the sequence $\left\{ \int_{S^n} K_n(\bx,\cdot,t) \r_n
(d\bx) \right\}$.

\end{definition}

 The quantum
analog of a Markov transition function is a completely positive unital
map.    A linear map $\phi: \A_1 \longrightarrow \A_2$ of C* algebras is {\it
completely positive} if, for each $n \in \NN$,
the map from $\A_1 \otimes \B(\CC^n)$ to $\A_2 \otimes \B(\CC^n)$ that
sends $A \otimes B$ to $\phi(A)\otimes B$ is positive
\cite{Davies}.
Propagation of chaos is an attribute of certain sequences of Markov
transition functions; we now define {\it quantum} propagation of chaos to be an
attribute of certain sequences of completely positive unital maps.

\begin{definition}[Propagation of Quantum Molecular Chaos]
\label{PropagationOfQuantumChaos}

For each $n \in \NN$, let $\phi_n$ be a completely positive map
from $\otimes^n \A$ to itself that fixes the unit ${\bf 1}\otimes
\cdots \otimes {\bf 1} \in \otimes^n \A$ and which commutes with
permutations, i.e., such that
\begin{equation}
\label{PermutationsPhi}
    \phi_n(A_{\pi(1)} \otimes A_{\pi(2)} \otimes \cdots \otimes
    A_{\pi(n)}) = \pi \cdot \phi_n (A_1 \otimes A_2 \otimes \cdots \otimes
    A_n)
\end{equation}
for all permutations $\pi$ of $\{1,2,\ldots,n\}$, where $\pi \cdot$
denotes the operator on $\otimes^n \A$ defined by
\[
   \pi \cdot (B_1 \otimes B_2 \otimes \cdots \otimes B_n) =
   B_{\pi(1)}\otimes B_{\pi(2)} \otimes \cdots \otimes B_{\pi(n)}
\]
for all $B_1,B_2,\ldots,B_n \in \A$.

The sequence $\{\phi_n\}$ {\bf propagates chaos} if the molecular
chaos of a sequence of states $\{ \r_n \}$ entails the molecular
chaos of the sequence $\{ \r_n \circ \phi_n \}$.

\end{definition}

Consider the case where $\A$ is the algebra generated by the observables
for a single
particle of a certain species, so that $\otimes^n \A$ is the
 algebra generated by all
single-particle observables in a system of $n$ {\it distinguishable}
particles of that species.
For each $n \in \NN$, let the dynamics of the
$n$-particle system be given by a Hamiltonian operator $H_n \in
\otimes^n \A$.  In the Heisenberg version of quantum dynamics, if $A$
is the operator that corresponds to measurement of a certain observable
quantity at $t=0$, the operator corresponding to the measurement of
the same quantity at time $t > 0$ equals $e^{i H_n t / \hbar } A e^{-i
H_n t / \hbar }$.
The maps $\phi_{n,t}:\otimes^n \A \rightarrow \otimes^n
\A$ defined  by
\begin{equation}
\label{Phi}
   \phi_{n,t}(A) =  e^{i H_n t / \hbar } A e^{-i H_n t /
                  \hbar }
\end{equation}
are completely positive, and if they satisfy condition
(\ref{PermutationsPhi}) one may ask whether the sequence $\{
\phi_{n,t} \}$ propagates chaos.

We conjecture that chaos always propagates when the
$n$-particle Hamiltonians $H_n$ are as follows:
Let $\A = \B(\CC^d)$, the algebra of all bounded operators on
$\CC^d$.  The algebra $\otimes^n \A$ is isomorphic to $\B(\otimes^n
\CC^d)$, and states $\tau \in \S(\otimes^n \A) $ correspond to density
operators $D_{\tau}$ on $\B(\otimes^n \CC^d)$ via the equation
$   \tau(A) = {\mathrm {Tr}}(D_{\tau}A)  $.
Suppose that $V$ is a Hermitian operator on $\CC^d \otimes \CC^d$ that is
symmetric in the sense that $V(x \otimes y) = V(y \otimes x)$ for all
$x,y \in \CC^d$.
Let $V^n_{1,2}$ denote the operator on $\otimes^n \CC^d$ defined by
\[
   V^n_{1,2}(x_1\otimes x_2 \otimes \cdots \otimes x_n) = V(x_1 \otimes
   x_2) \otimes x_3 \otimes \cdots \otimes x_n ,
\]
and for each $i < j \le n$, define $V^n_{ij}$ similarly  (as acting
on the $i^{th}$ and $j^{th}$ factors of each simple tensor).
Define the $n$-particle Hamiltonian $H_n$ as the sum of the pair
potentials $V^n_{ij}$, with a $1/n$ scaling of the
coupling constant:
\begin{equation}
\label{SpohnStyleHamiltonian}
   H_n = \oon \sum_{i < j} V^n_{ij}.
\end{equation}
\begin{conjecture}
\label{conjecture}

The sequence $\{ \phi_{n,t}\} $ defined in equation (\ref{Phi}) propagates
chaos:

 If $\{ \r_n \}$ is $\r$-chaotic then $\{ \r_n \circ \phi_{n,t} \}$ is
$\r_t$-chaotic, where the density operator for
$\r_t$ is the solution at time $t$ of
\begin{eqnarray}
\label{MeanFieldDynamics}
       \frac{\partial}{\partial t} D & = & -\frac{i}{\hbar}[V,D\otimes
D]^{(1)}
                                       \nonumber \\
       D(0) & = & D_{\r} .             \nonumber \\
\end{eqnarray}
Here $[V,D\otimes D]^{(1)}$ denotes a contraction of $[V,D\otimes
D]$: if $\{y_i\}$ is any orthonormal basis of $\CC^d$ and $x
\in \CC^d$,
\[
    [V,D\otimes D]^{(1)}(x) = \sum_{i} <[V,D\otimes D]
                (x \otimes y_i), (x \otimes y_i)>.
\]
\end{conjecture}

This conjecture will now be verified for
the Curie-Weiss model of ferromagnetism.
In this model, the ferromagnetic material is modelled by
a crystal in which the spin angular momentum of each atom is
coupled to the average spin and to an external magnetic field.  In case the
applied magnetic field is directed along the $z$-axis, we may
make the approximation that only the $z$-components of the spins are
coupled to
each other and the external field.
As in \cite{Emch, BagMor}, we consider the case of spin-$\frac{1}{2}$
atoms, so that the space of
pure spin states of a single particle is $\CC^2$, and the observables
corresponding to
the measurement of the $x, y$ and $z$ components of spin are the
Pauli spin operators
\[
\begin{array}{ccc}
   \sigma^x = \frac{\hbar}{2} \left( \begin{array}{cc} 0 & 1 \\ 1 & 0 \\
\end{array} \right)
   &
   \sigma^y = \frac{\hbar}{2} \left( \begin{array}{cc} 0 & -i \\ i & 0 \\
\end{array} \right)
   &
   \sigma^z = \frac{\hbar}{2} \left( \begin{array}{cc} 1 & 0 \\ 0 &
   -1 \\ \end{array} \right).
   \\
\end{array}
\]
The space of pure states of an $n$-spin system is $\otimes^n \CC^2$.
For each $i \le n$, if $A$ is an operator on $\CC$, let $A_i$ denote the
operator on $\otimes^n
\CC^2$ defined by
\[
    A_i(v_1 \otimes v_2 \otimes \cdots \otimes v_n) = v_1  \otimes
    \cdots \otimes A(v_i)  \otimes \cdots \otimes v_n   .
\]
If $A$ is Hermitian, $A_i$ corresponds to the measurement of the
spin observable $A$ at the $i^{th}$ lattice site.
The Hamiltonian for the $n$-site Curie-Weiss model is
\[
   \H_n = -J \sum_{i=1}^n \left( \sigma_i^z \frac
   {\sum_j\sigma_j^z}{n} \right) -
    H\sum_{i=1}^n \sigma_i^z = \oon \sum_{i,j =1}^n \left(
-J\sigma_i^z\sigma_j^z
   - H \sigma_i^z \right)  ,
\]
where $J$ is a positive coupling constant and $H$ is another constant
proportional to
the strength of the external magnetic field.

Since $\otimes^n \B(\CC^2)$ is isomorphic to $\B(\otimes^n \CC^2)$,
states $\r_n \in \S(\otimes^n \B(\CC^2))$ correspond to
density operators $D_{\r_n}$ in $\otimes^n \CC^2$.
>From definition (\ref{Phi}) and the
fact that ${\mathrm{Tr}}(AB) = {\mathrm{Tr}}(BA)$,
\begin{eqnarray*}
   \r_n \circ \phi_{n,t}(A)  & = & {\mathrm {Tr}}(D_{\r_n} e^{i\H_n
t/\hbar} A e^{-i
   \H_n t/\hbar}) \\ & = &  {\mathrm {Tr}}(e^{-i\H_n t/\hbar} D_{\r_n}
e^{i\H_n
   t/\hbar} A).
\end{eqnarray*}
Therefore, $\r_n \circ \phi_{n,t}$ has the density operator
\begin{equation}
\label{EvolvedDensity}
D_{\r_n \circ \phi_{n,t}} = e^{-i \H_n
t/\hbar} D_{\r_n} e^{i \H_n t/\hbar}.
\end{equation}

For any state $\r$ on $\B(\CC^2)$, let $D_{\r}$ denote the
corresponding density operator, and let $\left[ D_{\r(t)} \right]$
denote a $2 \times2$ matrix that represents $D_{\r}$.

\begin{theorem}
\label{CurieWeissTheorem}

The sequence of Hamiltonians $\{ \H_n \}$ propagates chaos.  If $\{
\r_n \}$ is a $\r$-chaotic sequence of states with $\left[
D_{\r}\right] =
\left( \begin{array}{cc} a & c \\ \bar{c} & d \\ \end{array}
\right)$, then for each $t \ge 0$,  $\{ \r_n \circ \phi_{n,t} \}$ is
$\r(t)$-chaotic, where
\[
\left[ D_{\r(t)} \right] =
\left( \begin{array}{cc} a & ce^{i t(H + \hbar J (a-d))} \\
           \bar{c}e^{- i t(H + \hbar J (a-d))} & d \\ \end{array}
\right).
\]

\end{theorem}

\noindent{\bf Proof of Theorem \ref{CurieWeissTheorem}}:

If $\bx \in \{ 0,1\}^n$ for some $n \in \NN$, let $\N(\bx)$ denote the
number of $1$s in $\bx$:
\[
   \N(\bx) = \sum_{i = 1}^n x_i
\]
if $\bx = (x_1,x_2,\ldots,x_n)$.
If $g:\{0,1\}^n \longrightarrow \CC$ is a symmetric function, then $g(\bx)$
depends on $\bx$ only through $\N(\bx)$, so that $g(\bx) = g(\by) $ if
$\N(\bx) = \N(\by)$.

\begin{lemma}
\label{CuteLemma}
For each $n \in \NN$, let $f_n:\{0,1\}^n \longrightarrow \CC$ be a
symmetric function.

 Suppose that
\begin{trivlist}
\item{(a)} there exists $B < \infty$ such that
   $\sum_{\bs \in \{0,1\}^n} \left|f_n(\bs)\right| \le B$ for all $n$,
   and
\item{(b)}
   there exists $c \in \CC$ and $f:\{0,1\}\longrightarrow \CC$ such
   that, for $k = 0,1,2,\ldots$,
\[
   \sum_{z_1,z_2,\ldots,z_{n-k}} f_n(x_1,\ldots,x_k,z_1,\ldots,z_{n-k})
   \longrightarrow cf(x_1)f(x_2)\cdots f(x_k)
\]
as $n \longrightarrow \infty$.
\end{trivlist}

Then
\begin{trivlist}
\item{(i)} For all $G \in C_{\CC}([0,1])$,
\[
   \lim_{n \rightarrow \infty} \sum_{\bx \in \{0,1\}^n}
   f_n(\bx) G \left(\frac{\N(\bx)}{n}\right) = c \  G(f(1))
\]
\item{(ii)}
If $ c\ne 0$ then   $0 \le f(0), f(1) \le 1$ and $f(0) + f(1) = 1$.
\end{trivlist}

\end{lemma}

\noindent{\bf Proof of Lemma}:
Let
\begin{equation}
\label{Fn}
   F_n(j) = \sum_{\N(\bx) = j} f_n(\bx).
\end{equation}
Condition (b) implies that
\[
   \lim_{n \rightarrow \infty} \sum_{z_1,z_2,\ldots,z_{n-k}}
f_n(1,1,\ldots,1,z_1,\ldots,z_{n-k})
   = c\left( f(1) \right)^k
\]
 for all $k \in \NN$.
Grouping the summands for which exactly $j$ coordinates
of $(1,1,\ldots,1,z_1,\ldots,z_{n-k})$ equal $1$, we find that
\begin{equation}
\label{Monomials}
   \lim_{n \rightarrow \infty} \sum_{j = k}^n  F_n(j)
   \frac{\left( \begin{array}{c} n -k \\ j - k \\ \end{array}\right)}
      {\left( \begin{array}{c} n \\ j \\ \end{array} \right) }
      = c \left( f(1) \right)^k  .
\end{equation}

Now
\[
   \sum_{j = k}^n  F_n(j)
    \frac{ \left( \begin{array}{c} n -k \\ j - k \\
    \end{array}\right)}
    { \left( \begin{array}{c} n \\ j \\ \end{array} \right) } =
    \sum_{j = k}^n  F_n(j)
   \frac{j (j-1)\cdots (j - k +1)}{n (n-1) \cdots (n-k+1)},
\]
while
\begin{eqnarray*}
&&
\left|
   \sum_{j = k}^n  F_n(j)
   \frac{j (j-1)\cdots (j - k +1)}{n (n-1) \cdots (n-k+1)}
   - \sum_{j = 0}^n  F_n(j)\left(\frac{j}{n}\right)^k
\right|    \\
& \le &
   \sum_{j = 0}^n \left| F_n(j) \right| \left( \left(
   \frac{k}{n}\right)^n + \max_{j \ge k}
   \left\{ \left|
   \frac{j (j-1)\cdots (j - k +1)}{n (n-1) \cdots (n-k+1)} -
   \left(\frac{j}{n}\right)^k \right| \right\} \right)  \\
& \le &
   B \left[ \left(\frac{k}{n}\right)^n + \max_{j \ge k}\left\{
     \left(\frac{j}{n}\right)^k - \left(\frac{j-k+1}{n}\right)^k
     \right\} \right] \\
& \le &
   B \left[ \left(\frac{k}{n}\right)^n + \frac{1}{n^k}\max_{j \ge k}\left\{
     \left( \begin{array}{c} k \\ 1 \\ \end{array}\right) j^{k-1}(k-1) +
     \left( \begin{array}{c} k \\ 2 \\ \end{array}\right)
     j^{k-2}(k-1)^2 + \cdots + (k-1)^k
     \right\} \right] \\
& \le &
   B \left[ \left(\frac{k}{n}\right)^n + \oon \left\{
     \left( \begin{array}{c} k \\ 1 \\ \end{array}\right) k +
     \left( \begin{array}{c} k \\ 2 \\ \end{array}\right) k^2 + \cdots + k^k
     \right\} \right] \\
& \le &
  B  \left[ \left(\frac{k}{n}\right)^n + \oon(1+k)^k \right],
\\
\end{eqnarray*}
and therefore
\begin{equation}
\label{Asymptotic}
   \lim_{n \rightarrow \infty} \sum_{j = k}^n  F_n(j)
   \frac{\left( \begin{array}{c} n -k \\ j - k \\ \end{array}\right)}
      {\left( \begin{array}{c} n \\ j \\ \end{array} \right) } =
   \lim_{n \rightarrow \infty} \sum_{j = 0}^n  F_n(j)\left(
   \frac{j}{n}\right)^k.
\end{equation}

Condition (a) implies that $\sum_j \left| F_n(j)
\right| \le B$ for all $n$.  This bound, equations (\ref{Monomials})
and (\ref{Asymptotic}), and the fact that any
continuous function on $[0,1]$ can be approximated uniformly by
polynomials, imply that
\[
   \lim_{n \rightarrow \infty} \sum_{j = 0}^n  F_n(j) G\left(
   \frac{j}{n}\right)   = c \  G\left( f(1) \right)
\]
for all $G \in C_{\CC}([0,1])$.
This establishes (i), in view of the definition (\ref{Fn}) of $F_n$.

If $c \ne 0$, conclusion (i) implies that $f(1) \in [0,1]$.
Condition (b) for $k = 0$ and $k = 1$ states that
\begin{eqnarray*}
     \lim_{n \rightarrow \infty} \sum_{z_1,z_2,\ldots,z_n}
     f_n(z_1,z_2,\ldots,z_n)
   & = &   c  \\
     \lim_{n \rightarrow \infty} \sum_{z_1,z_2,\ldots,z_{n-1}}
f_n(1,z_1,\ldots,z_{n-1})
   & = &   c f(1)    \\
   \lim_{n \rightarrow \infty} \sum_{z_1,z_2,\ldots,z_{n-1}}
f_n(0,z_1,\ldots,z_{n-1})
   & = &   c f(0) ,   \\
\end{eqnarray*}
 so it follows that $f(0)+f(1)=1$ if $c \ne 0$.  This establishes
(ii), concluding the proof of the lemma.

\noindent $\blacksquare$

Let $e_1 = (1,0)$ and $e_2 = (0,1)$.  For each $n \in \NN$, let
\[
   \E_n = \left\{ e_{j_1} \otimes e_{j_2} \otimes \cdots \otimes e_{j_n}
   \in \otimes^n \CC^2| \  j_1,j_2,\ldots,j_n \in \{1,2\} \right\}.
\]
Denote by $[A]_{j_1,\ldots,j_n}^{ k_1,\ldots,k_n}$ the matrix elements for
an operator
$A$ on $\otimes^n \CC^2$ relative to the basis
$\E_n$, that
is, for
$j_1,k_1,\ldots,j_n,k_n \in \{1,2\}$, let
\[
     [A]_{j_1,\ldots,j_n}^{ k_1,\ldots,k_n} =
         \left< A(e_{j_1} \otimes e_{j_2} \otimes \cdots \otimes
         e_{j_n}), e_{k_1} \otimes e_{k_2} \otimes \cdots \otimes
         e_{k_n} \right>.
\]

The sequence $\{ \r_n \}$ is $\r$-chaotic.  This implies that, for
each $k \in \NN$,
\[
    \sum_{z_1,z_2,\ldots,z_{n-k} \in \{1,2\}} \left[ D_{\r_n} \right]
       _{x_1,\ldots,x_k,z_1,\ldots,z_{n-k}}^{
       y_1,\ldots,y_k,z_1,\ldots,z_{n-k}}
     \longrightarrow
         \left[D_{\r}\right]_{x_1}^{y_1}\left[D_{\r}\right]_{x_2}^{y_2}
          \cdots \left[D_{\r}\right]_{x_k}^{y_k} ,
\]
as $n \longrightarrow \infty$.   To show that $\{ \r_n \circ
\phi_{n,t} \}$ is $\r(t)$-chaotic, it suffices to show that for all
$k \in \NN$,
\begin{equation}
\label{Sufficient}
    \sum_{z_1,z_2,\ldots,z_{n-k} \in \{1,2\}} \left[ D_{\r_n \circ
    \phi_{n,t}} \right]
       _{x_1,\ldots,x_k,z_1,\ldots,z_{n-k}}^{
       y_1,\ldots,y_k,z_1,\ldots,z_{n-k}}
     \longrightarrow
         \left[D_{\r(t)}\right]_{x_1}^{y_1}\left[D_{\r(t)}\right]_{x_2}^{y_2}
          \cdots \left[D_{\r(t)}\right]_{x_k}^{y_k}
\end{equation}
as $n \longrightarrow \infty$.  We proceed to verify
(\ref{Sufficient}).

The operator $\H_n$ is diagonalized by
the basis $\E_n$, and its diagonal entries are
\begin{equation}
\label{DiagonalH}
        [H_n]_{k_1,\ldots,k_n}^{ k_1,\ldots,k_n} =  \oon
        \sum_{r,s=1}^n \left(-J\eta(e_{k_r})\eta(e_{k_s})
-H\eta(e_{k_r})\right),
\end{equation}
where
\begin{eqnarray*}
                      \eta(e_1) & = & +\frac{\hbar}{2}    \\
                      \eta(e_2) & = & -\frac{\hbar}{2} .   \\
\end{eqnarray*}
Abbreviating $-J\eta(x)\eta(y) -H\eta(x)$ by $w(x,y)$,
equations (\ref{EvolvedDensity}) and (\ref{DiagonalH}) imply that
\begin{eqnarray*}
&&
     [D_{\r_n \circ \phi_{n,t}}]
       _{x_1,\ldots,x_k,z_1,\ldots,z_{n-k}}^{
y_1,\ldots,y_k,z_1,\ldots,z_{n-k}} = \\
&&
     \exp\left( \frac{it}{\hbar n} \sum_{r,s=1}^k
     \left( w(y_r,y_s) - w(x_r,x_s) \right) \right)
     \exp\left( H\frac{it}{\hbar}\frac{n-k}{ n} \sum_{r=1}^k
     (\eta(x_r) - \eta(y_r) ) \right)  \times   \\
&&
     \left[ D \right]_{x_1,\ldots,x_k,z_1,\ldots,z_{n-k}}^{
y_1,\ldots,y_k,z_1,\ldots,z_{n-k}}
     \exp\left( J\frac{2it}{\hbar n} \sum_{r=1}^k \sum_{s=k+1}^n
       \eta(z_s)(\eta(x_r) - \eta(y_r)) \right).           \\
\end{eqnarray*}
Therefore,
\begin{eqnarray*}
  &&
   \lim_{n \rightarrow \infty}\sum_{z_1,\ldots,z_{n-k} \in \{1,2\}} \left[
D_{\r_n \circ
    \phi_{n,t}} \right]_{x_1,\ldots,x_k,z_1,\ldots,z_{n-k}}^{
    y_1,\ldots,y_k,z_1,\ldots,z_{n-k}} \\
 & = &
    \exp\left( H\frac{it}{\hbar} \sum_{r=1}^k
     (\eta(x_r) - \eta(y_r) ) \right)
     \times
    \\
 &&
    \lim_{n \rightarrow \infty}\sum_{z_1,\ldots,z_{n-k} \in \{0,1\}}
     \left[ D \right]_{x_1,\ldots,x_k,z_1,\ldots,z_{n-k}}^{
y_1,\ldots,y_k,z_1,\ldots,z_{n-k}}
     \exp\left( J\frac{2it}{\hbar n} \sum_{r=1}^k \sum_{s=k+1}^n
      \eta(z_s) (\eta(x_r) - \eta(y_r)) \right)          \\
 & = &
    \exp\left( H\frac{it}{\hbar} \sum_{r=1}^k
     (\eta(x_r) - \eta(y_r) ) \right)
     \times
    \\
 &&
     \lim_{n \rightarrow \infty}\sum_{z_1,\ldots,z_{n-k} \in \{1,2\}}
     \left[ D \right]_{x_1,\ldots,x_k,z_1,\ldots,z_{n-k}}^{
y_1,\ldots,y_k,z_1,\ldots,z_{n-k}}
     \exp\left( J\frac{2it}{\hbar}\sum_{r=1}^k (\eta(x_r) - \eta(y_r))
        \oon \sum_{s=k+1}^n \eta(z_s) \right)     .     \\
\end{eqnarray*}

The last limit in the preceding equation may be calculated thanks to
Lemma~\ref{CuteLemma}.
To apply the lemma, fix $x_1,y_1,\ldots,x_k,y_k \in \{ 1,2 \}$ and
define
\begin{eqnarray*}
    f_n: \{1,2 \} \longrightarrow \CC;  &&
    f_n(z_1,z_2,\ldots,z_n) =
    \left[ D \right]_{x_1,\ldots,x_k,z_1,\ldots,z_n}
                    ^{y_1,\ldots,y_k,z_1,\ldots,z_n} \\
    G:[0,1] \longrightarrow \CC; &&
    G(s) = \exp \left( J i t \sum_{r=1}^k (\eta(x_r) - \eta(y_r))
            (s - (1 - s)) \right) . \\
\end{eqnarray*}
The functions $f_n$ are symmetric and satisfy condition (b) of the
lemma, with
\begin{eqnarray*}
   f(1) & = & \left[ D_{\r}\right]_{e_1}^{e_1}     \\
   f(2) & = & \left[ D_{\r}\right]_{e_2}^{e_2}         \\
      c & = & \left[D_{\r}\right]_{x_1}^{y_1}\left[D_{\r}\right]_{x_2}^{y_2}
          \cdots \left[D_{\r}\right]_{x_k}^{y_k} .  \\
\end{eqnarray*}
The $f_n$ also satisfy condition (a) of the lemma, for
\[
   \left| \left[ D \right]_{x_1,\ldots,x_k,z_1,\ldots,z_n}
                    ^{y_1,\ldots,y_k,z_1,\ldots,z_n} \right| \le
   \frac{1}{2}\left(
     \left[ D \right]_{x_1,\ldots,x_k,z_1,\ldots,z_n}
                    ^{x_1,\ldots,x_k,z_1,\ldots,z_n} +
     \left[ D \right]_{y_1,\ldots,y_k,z_1,\ldots,z_n}
                    ^{y_1,\ldots,y_k,z_1,\ldots,z_n}  \right)
\]
by the positivity of $D_{\r_n}$, so that
\[
    \sum_{z_1,\ldots,z_{n-k} \in \{1,2\}}
             \left| \left[ D \right]_{x_1,\ldots,x_k,z_1,\ldots,z_n}
                                    ^{y_1,\ldots,y_k,z_1,\ldots,z_n} \right|
     \le {\mathrm {tr}}(D_{\r_n})  = 1.
\]

Conclusion (i) of Lemma~\ref{CuteLemma} now reveals that
\[
  \sum_{z_1,\ldots,z_{n-k} \in \{1,2\}}
     \left[ D \right]_{x_1,\ldots,x_k,z_1,\ldots,z_{n-k}}^{
y_1,\ldots,y_k,z_1,\ldots,z_{n-k}}
     \exp\left( J\frac{2it}{\hbar} \sum_{r=1}^k (\eta(x_r) - \eta(y_r))
      \oon \sum_{s=k+1}^n \eta(z_s) \right)
\]
converges to
\[
  \left[D_{\r}\right]_{x_1}^{y_1}\left[D_{\r}\right]_{x_2}^{y_2}
          \cdots \left[D_{\r}\right]_{x_k}^{y_k}
          \exp\left( J i t \left( \left[ D_{\r}\right]_{e_1}^{e_1}  -
         \left[ D_{\r}\right]_{e_2}^{e_2} \right)
         \sum_{r=1}^k (\eta(x_r) - \eta(y_r))  \right)  ,
\]
whence
\begin{eqnarray*}
  &&
   \lim_{n \rightarrow \infty}\sum_{z_1,\ldots,z_{n-k} \in \{1,2\}} \left[
D_{\r_n \circ
    \phi_{n,t}} \right]_{x_1,\ldots,x_k,z_1,\ldots,z_{n-k}}^{
    y_1,\ldots,y_k,z_1,\ldots,z_{n-k}} \\
 & = &
     \exp\left( H\frac{it}{\hbar} \sum_{r=1}^k
     (\eta(x_r) - \eta(y_r) ) \right)
     \times
    \\
 &&
  \left[D_{\r}\right]_{x_1}^{y_1}\left[D_{\r}\right]_{x_2}^{y_2}
          \cdots \left[D_{\r}\right]_{x_k}^{y_k}
          \exp\left( J i t \left( \left[ D_{\r}\right]_{e_1}^{e_1}  -
         \left[ D_{\r}\right]_{e_2}^{e_2} \right)
         \sum_{r=1}^k (\eta(x_r) - \eta(y_r))  \right)  \\
 & = &
    \prod_{r = 1}^k \left[D_{\r}\right]_{x_r}^{y_r}
     \exp\left\{ i t \left( \eta(x_r) - \eta(y_r) \right)
    \left( H /\hbar
    + J \left(\left[ D_{\r}\right]_{e_1}^{e_1}  -
                 \left[ D_{\r}\right]_{e_2}^{e_2} \right) \right)\right\}  .
\\
\end{eqnarray*}

This shows that $\{ \r_n \circ \phi_{n,t} \}$ is $\r(t)$-chaotic,
where
\[
   \left[ D_{\r(t)} \right]_x^y =  \left[D_{\r}\right]_x^y
     \exp\left\{ i t \left( \eta(x) - \eta(y) \right)
    \left( H /\hbar
    + J \left(\left[ D_{\r}\right]_{e_1}^{e_1}  -
                 \left[ D_{\r}\right]_{e_2}^{e_2} \right)\right) \right\} .
\]

Finally,
it may be verified that the density operators $D_{\r(t)}$ satisfy
equation~(\ref{MeanFieldDynamics}) of Conjecture~\ref{conjecture}.

\noindent $\blacksquare$

\section{Future Work}

In future work, we hope to prove Conjecture~\ref{conjecture},
or at least to prove that the mean-field Heisenberg
model propagates chaos.  We shall also investigate the propagation of chaos
by open
systems (coupled to thermal baths) and prove the H-theorem for those
processes.

\section{Acknowledgements}

I am indebted to William Arveson, Sante Gnerre, Lucien
Le Cam,  Marc Rieffel, and Geoffrey Sewell for their advice and for
their interest in this research.   The formulation and proof of
Theorem~\ref{Elizabeth} was especially inspired by Elizabeth Kallman.
I thank Alexandre Chorin for introducing me to statistical physics and
other topics in applied mathematics.
This work was supported in part by the
Office of Science, Office of
Advanced Scientific Computing Research,
Mathematical, Information, and Computational Sciences Division,
Applied Mathematical Sciences Subprogram, of
the U.S.\ Department of Energy, under Contract No.\ DE-AC03-76SF00098.

\end{document}